\documentclass[pra,twocolumn,showpacs,superscriptaddress,amssymb,10pt]{revtex4-1}

\usepackage{graphicx}
\usepackage{dcolumn}
\usepackage{bm}
\usepackage{epsfig}
\usepackage{color}
\usepackage{longtable}
\usepackage{amsmath}
\usepackage{multirow}
\usepackage{tabularx}
\usepackage{siunitx}
\definecolor{aogreen}{rgb}{0.0, 0.5, 0.0}

\def\ketm#1{  \left\vert  #1   \right\rangle   }

\def\sprm#1#2{  \left\langle #1 \left\vert \right. #2 \right\rangle   }

\definecolor{mymainmessagecolor}{RGB}{10,200,10}
\definecolor{revisedcolor}{RGB}{0,100,20}
\def\Rolll{U_{p_{1/2}}^{(s_{1/2})}}
\def\Rdtll{U_{p_{1/2}}^{(d_{3/2})}}
\def\Rollt{U_{p_{3/2}}^{(s_{1/2})}}
\def\Rdtlt{U_{p_{3/2}}^{(d_{3/2})}}
\def\Rdflt{U_{p_{3/2}}^{(d_{5/2})}}
\def\Rdttf{U_{f_{5/2}}^{(d_{3/2})}}
\def\Rdftf{U_{f_{5/2}}^{(d_{5/2})}}
\def\Rdfts{U_{f_{7/2}}^{(d_{5/2})}}

\DeclareMathOperator*{\SumInt}{%
\mathchoice%
  {\ooalign{$\displaystyle\sum$\cr\hidewidth$\displaystyle\int$\hidewidth\cr}}
  {\ooalign{\raisebox{.14\height}{\scalebox{.7}{$\textstyle\sum$}}\cr\hidewidth$\textstyle\int$\hidewidth\cr}}
  {\ooalign{\raisebox{.2\height}{\scalebox{.6}{$\scriptstyle\sum$}}\cr$\scriptstyle\int$\cr}}
  {\ooalign{\raisebox{.2\height}{\scalebox{.6}{$\scriptstyle\sum$}}\cr$\scriptstyle\int$\cr}}
}

\begin{document}
\include{Bibliography.bib}
\preprint{}
\title{
Enhanced polarization transfer to the characteristic $L\alpha$ x-ray lines near the nonlinear Cooper minimum of two-photon ionization
}

\author{J.~Hofbrucker}
\affiliation{Helmholtz-Institut Jena, Fr\"o{}belstieg 3, D-07743 Jena, Germany}
\affiliation{Theoretisch-Physikalisches Institut, Friedrich-Schiller-Universit\"at Jena, Max-Wien-Platz 1, D-07743
Jena, Germany}%
\affiliation{GSI Helmholtzzentrum f\"ur Schwerionenforschung GmbH, Planckstrasse 1, D-64291 Darmstadt, Germany}

\author{A.~V.~Volotka}
\affiliation{Helmholtz-Institut Jena, Fr\"o{}belstieg 3, D-07743 Jena, Germany}%
\affiliation{GSI Helmholtzzentrum f\"ur Schwerionenforschung GmbH, Planckstrasse 1, D-64291 Darmstadt, Germany}

\author{J.~ Szlachetko}
\affiliation{Institute of Nuclear Physics, Polish Academy of Sciences, ul. Radzikowskiego 152, PL-31342 Krakow, Poland}%

\author{S.~Fritzsche}
\affiliation{Helmholtz-Institut Jena, Fr\"o{}belstieg 3, D-07743 Jena, Germany}%
\affiliation{Theoretisch-Physikalisches Institut, Friedrich-Schiller-Universit\"at Jena, Max-Wien-Platz 1, D-07743
Jena, Germany}
\affiliation{GSI Helmholtzzentrum f\"ur Schwerionenforschung GmbH, Planckstrasse 1, D-64291 Darmstadt, Germany}

\date{\today \\[0.3cm]}

\begin{abstract}
It has been shown that for nonlinear atomic ionization, transfer of the degree of circular polarization from incident to fluorescence light is maximum at the Cooper minimum (see Phys. Rev. A \textbf{100}, 011401(R) (2019)). Until the present, however, it is still a challenge to produce and detect circularly-polarized light at xuv and x-ray photon energies. We here show, that transfer of linear polarization is strongly enhanced at the Cooper minimum in two-photon ionization and can be readily detected using current free-electron laser facilities by measurements of the degree of linear polarization of the characteristic $L\alpha_1$ and $L\alpha_2$ lines. Two-photon ionization of $2p_{3/2}$ electron of tungsten atom is proposed to demonstrate the presented phenomena and realization of such experiment is discussed in detail.
\end{abstract}

\newpage
\maketitle


\section{Introduction}
\label{Sec.Intro}

First experimental observations of two-photon absorption mechanism involving core-level atomic states have become possible only rather recently by virtue of the development of x-ray Free Electron Lasers (XFELs) \cite{Emma:2010jl, Amann:2012cp}. Since the cross-sections of two-photon absorption and ionization process are generally low, it is often necessary to employ high brilliance incident x-ray beams with properties such as coherence, duration of the pulse and intensity, which are otherwise known only from optical lasers. With the advent of XFEL facilities, x-ray beams with femtosecond pulse duration and photon flux of $10^{31}$~photons/(s cm$^{2})$ become achievable. The application of such high-power x-ray beams enabled one to approve nonlinear processes with quite low excitation probabilities, including multiple ionization \cite{Young:2010:56, Vinko:2012gx}, double-core hole creation \cite{2013PhRvL.111d3001T} and nonsequential two-photon ionization \cite{Richardson:2010:013001, Doumy:2011:083002}. More recently, experimental observations of two-photon ionization of $K$-shell electrons at hard x-ray energies were reported \cite{Tamasaku:2014:313, Ghimire:2016:043418, Szlachetko:2016:33292, Tamasaku:2018:083901, Kayser:2019ky}. 

Core-hole excitation and ionization processes can be effectively investigated by observing the subsequent fluorescence photons. In particular, the photon polarization carries information about the alignment and orientation of an atom, which, in turn, strongly depends on details of the core electron ionization or excitation \cite{Fano:1973:553, Klar:1980:2037, Starace:1982:842, Huang:1982:3438}. The polarization properties of fluorescence light following single photon ionization is a well developed part of atomic spectroscopy \cite{Schmidt:1992:1483, Kabachnik:2007:155, Sharma:2010:023419, Sukhorukov:2019:1} in contrast to the nonlinear ionization, which is much less explored. In recent experiments, the total two-photon inner-shell ionization cross sections have been extracted from the subsequent fluorescence yields \cite{Tamasaku:2014:313, Ghimire:2016:043418, Szlachetko:2016:33292, Tamasaku:2018:083901, Kayser:2019ky}. However, these measurements provide only the total two-photon ionization cross section, which generally suffers from low precision. Performing similar experiments together with detection of the polarization of the fluorescence photons would allow to investigate the magnetic population dynamics in the produced ion and reach higher precision of the extracted ionization pathways (channels) contributions.

In one-photon ionization of atoms, the energy at which the dominant ionization channel vanishes is called the Cooper minimum \cite{Cooper:1964:762}. It has been shown both theoretically and experimentally, that performing measurements at the Cooper minimum can reveal otherwise hidden contributions to the ionization process, such as the contributions of higher multipole orders of the electromagnetic field \cite{Dehmer:1976:1049, Johnson:1978:1167, White:1979:1661, Grum:2001:L359, Grum:2014:043424, Ilchen:2018:4659}. Cooper minimum in multi-photon ionization can be also noticed in earlier calculations, both in total cross sections \cite{Lambropoulos:1987:821, Saenz:1999:5629, Nikolopoulos:2006:043408} as well as in photoelectron angular distributions \cite{Lagutin:2017:063414, Petrov:2019:013408}, however, no attention was paid to investigate it in detail. We recently demonstrated, that such Cooper minimum in multi-photon ionization processes can strongly influence the photoelectron angular distributions \cite{Hofbrucker:2018:053401}, or the magnetic sublevel populations of photoions \cite{Hofbrucker:2019:011401, Hofbrucker:2020:3617}.  In \cite{Hofbrucker:2019:011401}, it has also been shown, that in the case of two-photon ionization of atomic $p$ electrons by circularly polarized light, the polarization transfer from the ionizing light to the atom is maximized at this \textit{nonlinear Cooper minimum}. However, although it is possible to generate intense circularly polarized beams at some free-electron laser facilities \cite{Allaria:2013:913, Lutman:2016:468}, the production of such laser beams are usually more limited in photon energy than linearly polarized light. Moreover, detection of polarization of XUV or x-ray is experimentally very challenging. On the other hand, linearly polarized light can be produced at practically all free-electron facilities, and moreover, detection of the degree of linear polarization in x-ray domain can be carried out with high accuracy \cite{Marx:2013.254801}.

In this paper, we study the transfer of linear polarization in the two-photon ionization of atomic $p_{3/2}$ electrons. In the next section, we describe our theoretical approach based on the second-order perturbation and the density matrix theories as well as independent-particle approximation. In Sec. \ref{Sec.Results}, we present our calculations which were chosen to fulfill current experimental possibilities. On the example, we demonstrate the dynamic behavior of the polarization transfer in two-photon ionization of $2p_{3/2}$ electron of tungsten as a function of incident photon energy. Moreover, we show that significant variation of the polarization transfer occurs at the nonlinear Cooper minimum, whilst the corresponding cross sections are comparable to the already performed experiments \cite{Tamasaku:2014:313, Ghimire:2016:043418, Szlachetko:2016:33292, Tamasaku:2018:083901}. A discussion of the possible experimental possibilities to perform the suggested experiment are provided in Sec. \ref{Sec.Experiment}. In the last section, we conclude the main findings of this paper.


%
\section{Theory}
\label{Sec.Theory}

\begin{figure}
    \includegraphics[scale=0.5]{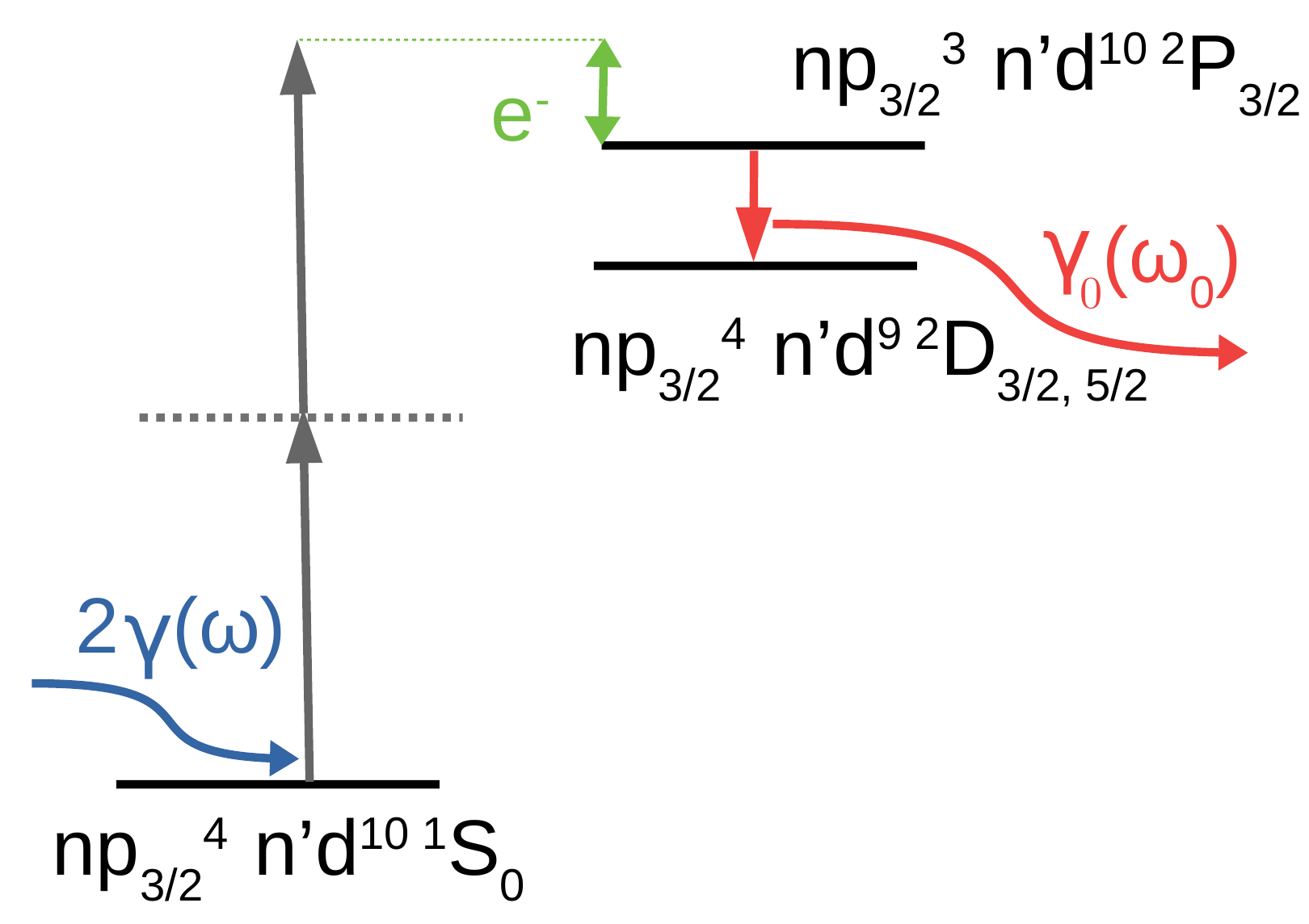}
    \caption{Schematic representation of the two-photon ionization process with subsequent fluorescence decay. One of the $n p^4_{3/2}$ electrons of an atom A absorbs two linearly polarized photons $\gamma (\omega)$ and is ionized to continuum state $e^-$. The produced photoion A$^+$ decays to a lower energy state with a hole in a $n' d$ state by emitting a fluorescence photon $\gamma_0(\omega_0)$. The dashed horizontal line represents an intermediate state which can be either real or virtual. Both resonant and nonresonant two-photon ionization are considered in this work.}
\label{Fig.Process} 
\end{figure}

We consider a two-step process. In the first step, an inner-shell ionization of an atom in an initial state $\ketm{\alpha_i J_i M_i}$ by two identical, linearly polarized photons $\gamma (\omega)$ with energy $\omega$ results in emission of a photoelectron with well-defined momentum $\bm{p}_e $ and spin projection $m_e$, $\ketm{\bm{p}_e m_e}$ and production of an ion in a state $\ketm{\alpha_f J_f M_f}$, which has typically a hole in one of the core-shells. Here, the many-electron wave functions of the atom are classified by the total angular momentum $J$, its projection $M$ as well as all further quantum numbers
$\alpha$ that are needed to specify the state uniquely. In the second step, the excited ion $\ketm{\alpha_f J_f M_f}$ decays into a lower energy state $\ketm{\alpha_0 J_0 M_0}$ with spontaneous emission of a fluorescence photon $\gamma_0(\omega_0)$. This two-step process can be schematically represented as (see also Fig. \ref{Fig.Process})
\begin{eqnarray}
    \label{Eq.ProcessSchematics}
    \ketm{\alpha_i J_i M_i} + 2 \gamma (\omega) &\rightarrow& \ketm{\alpha_f J_f M_f} + \ketm{\bm{p}_e m_e} \rightarrow \\\nonumber
    &\rightarrow& \ketm{\alpha_0 J_0 M_0} + \ketm{\bm{p}_e m_e} + \gamma_0(\omega_0).
\end{eqnarray}
In this paper, we will only consider the two-photon ionization of an electron from the (filled) $2p_{3/2}^4$ shell of an atom with total angular momentum $J_i=0$, which leads to a photoion with the total angular momentum $J_f=3/2$. In the independent-particle model and by using second-order perturbation theory, the (two-photon) ionization can then be described by single-electron amplitudes \cite{Hofbrucker:2016:063412, Hofbrucker:2017:013409}, including the summation over the complete (single-) electron spectrum of the virtual intermediate states, with symmetries determined by the orbital and total angular momenta $l_n$ and $j_n$. The final photoelectron wave function $\ketm{\bm{p}_e m_e}$ is expanded into partial waves, each with the angular momenta $l$ and $j$. This expansion allows us to characterize the process in terms of angular momentum ionization channels. In the electric dipole approximation, the ionization of the $2p_{3/2}$ electron proceeds through intermediate states with $s$ and $d$ symmetries, into final photoelectron $p$ and $f$ states. There are eight possible relativistic ionization channels, each described by an amplitude $U_{l_{j}}^{(l_{n_{j_n}})}$. Once, the photoelectron has left the ion, all further (decay) properties of the photoion can be described by its density matrix $\rho_{M_f M_f'}$. In practice, however, it is often more convenient to characterize the ion polarization in terms of (so-called) statistical tensors $\rho_{kq}$ that transform like spherical tensor of rank $k$. These tensors are given by

\begin{equation}
    \rho_{kq} = \sum_{M_f M_f'} (-1)^{J_f-M_f'} \sprm{J_f M_f, J_f M_f'}{k q} \rho_{M_f M_f'}.
\end{equation}
For the two-photon ionization of a $p_{3/2}$ electron from a closed shell atom, the statistical tensor~$\rho_{kq}$ takes the form (see \cite{Hofbrucker:2019:011401, Hofbrucker:2020:3617} for details)
\begin{eqnarray}
    \label{Eq.StatisticalTensors}
    \rho_{20}&=& \Big[ 245 U_{p_{1/2}}^2 + 98 U_{p_{3/2}}^2 -27 U_{f_{5/2}}^2+6480 U_{f_{7/2}}^2 \Big], \nonumber\\
    \rho_{2 \pm2}&=& -\frac{\sqrt{3}}{\sqrt{2}} \rho_{20},\\
        \rho_{00}&=& 70\Big[ 7 U_{p_{1/2}}^2 + \frac{7}{25} \Bar{U}_{p_{3/2}}^{2} + \frac{27}{25} U_{f_{5/2}}^2 + 648 U_{f_{7/2}}^2 \Big], \nonumber\\ \nonumber 
        \rho_{2 \pm1} &=& \rho_{1 q} = \rho_{3 q} = 0, ~ \textrm{for all } q,
\end{eqnarray}
where the exact form of the transition amplitudes $U_{l_{j}}$ in terms of $U_{l_{j}}^{(l_{n_{j_n}})}$ is given in the Appendix \ref{Sec.Appendix}. The calculations presented in this work have been also carried out with contributions of higher multipole orders. However, the contributions of higher multipole orders were found negligible, and hence, electric dipole approximation is well justified. In contrast, the results presented in Ref. \cite{Hofbrucker:2020:3617} concentrate on a special case of two-photon ionization of $s$ electrons by circularly polarized light near the nonlinear Cooper minimum. In this particular scenario, all electric dipole channels drop to zero, and hence, reveal the contributions of higher multipole orders. The total two-photon ionization cross section $\sigma$ can be easily obtained from the zero rank statistical tensor $\rho_{00}$ as $\sigma(\omega) = 64 \alpha^2 \pi^5/\omega^2$ $\rho_{00}$. Moreover, statistical tensors are often utilized also to characterize the magnetic state of an ion. For example, an atom is said to be \textit{unpolarized} if it is described by just a single nonzero (statistical) scalar $\rho_{00}(J)$, while the atom or ion is called \textit{polarized} if at least one nontrivial ($k > 0$) statistical tensor occurs. Moreover, if only tensors or even rank occurs in the representation of the ion, it is called \textit{aligned}. Finally, if at least one odd rank statistical tensor is nonzero, the system is called \textit{oriented}. It is well-known, that for ionization of atom by linearly polarized light, the excited atom can be aligned along the direction of the photon momentum, but cannot be oriented. As seen from the expressions (\ref{Eq.StatisticalTensors}), the two-photon ionization by linearly polarized light leads to photoions that are described by even rank tensors only. On the contrary, two-photon ionization by circularly polarized photons produces aligned as well as oriented ion states, which can even reach pure orientation \cite{Hofbrucker:2019:011401}. To quantify the ion polarization, it is convenient to introduce the normalized statistical tensor $\mathcal{A}_{kq}$ defined as 
\begin{eqnarray}
    \label{Eq.AlignmentParameter}
    \mathcal{A}_{kq}&=& \frac{\rho_{kq}}{\rho_{00}}.
\end{eqnarray}
The component $\mathcal{A}_{20}$ is called the alignment parameter (or simply alignment) and expresses the difference in population between electrons with projections $M_f=\pm1/2$ and $\pm3/2$. This alignment parameter depends of course on the prior excitation or ionization process, and is fully described by the transition amplitudes $U_{l_{j}}^{(l_{n_{j_n}})}$ for the two-photon ionization of an initially closed-shell atom.

The alignment of an excited ion propagates also to the characteristics of subsequently emitted photons. We here consider the emission of the characteristic $L\alpha_1$ and $L\alpha_2$ lines, which arise from the $3d_{5/2} \rightarrow 2p_{3/2}$ and $3d_{3/2} \rightarrow 2p_{3/2}$ transitions, respectively. From the statistical tensors, we can easily obtain the general expression for the degree of linear polarization $P_l^{(J_0)}$ 
\begin{eqnarray}
    \label{Eq.StokesGeneral}
    P_l^{(d_{3/2})}&=& \frac{ 2\sqrt{6}\mathcal{A}_{22}+\sin^2 \theta (3\mathcal{A}_{20}-\sqrt{6}\mathcal{A}_{22}) }{5-2\mathcal{A}_{20}+\sin^2\theta(3\mathcal{A}_{20}-\sqrt{6}\mathcal{A}_{22})},\nonumber \\
    P_l^{(d_{5/2})}&=& \frac{ -2\sqrt{6}\mathcal{A}_{22}-\sin^2 \theta (3\mathcal{A}_{20}-\sqrt{6}\mathcal{A}_{22}) }{20+2\mathcal{A}_{20}-\sin^2\theta(3\mathcal{A}_{20}-\sqrt{6}\mathcal{A}_{22})}, \nonumber \\
\end{eqnarray}
where $\theta$ specifies the fluorescence photon emission direction with respect to the incident beam propagation direction and where we used the fact that $\mathcal{A}_{22} = \mathcal{A}_{2-2}$. We can further use the constant factor relation between $\mathcal{A}_{20}$ and $\mathcal{A}_{22}$ from Eqs. (\ref{Eq.StatisticalTensors}), which remains valid only for the case of two-photon ionization of $2p_{3/2}$ electrons by linearly polarized light. The relation of the two statistical tensors together with setting $\theta = 0$ allows us to obtain corresponding simplified expression for the degree of linear polarization of the $L\alpha$ lines
\begin{eqnarray}
    \label{Eq.StokesSpecific}
    P_l^{(d_{3/2})}&=& \frac{-6\mathcal{A}_{20}}{5-2\mathcal{A}_{20}}\nonumber \\
    P_l^{(d_{5/2})}&=& \frac{3\mathcal{A}_{20}}{10+\mathcal{A}_{20}}.
\end{eqnarray}
This simple expression reveals that $P_l^{(d_{3/2})}$ is generally larger than $P_l^{(d_{5/2})}$, and hence, experimental detection of the $L\alpha_1$ characteristic will yield a stronger signal. The above expressions also confirm the physical intuition that the degree of linear polarization of the emitted photon increases with the degree of ion alignment. For the process under consideration, $A_{20}\leq 1/2$. The maximum alignment $A_{20} = 1/2$ is reached when the contributions of the $U_{p_{1/2}}$ amplitude strongly dominates over others. In the case of maximum value of the alignment, the polarization of $L \alpha_2$ fluorescence reaches $P_l^{(d_{3/2})} = 75\%$ at $\theta = 0$. 

More importantly, Eq. (\ref{Eq.StokesSpecific}) shows that the polarization of the $L\alpha$ lines is determined by a single variable $\mathcal{A}_{20}$. Since the alignment expresses the ratio of probabilities of creation of $|M_f|=1/2$ and $3/2$ holes, any variation in the degree of polarization of the $L\alpha$ lines directly expresses the ratio of populations of the corresponding substates.

\begin{figure*}
    \centering
    \includegraphics[width=\textwidth]{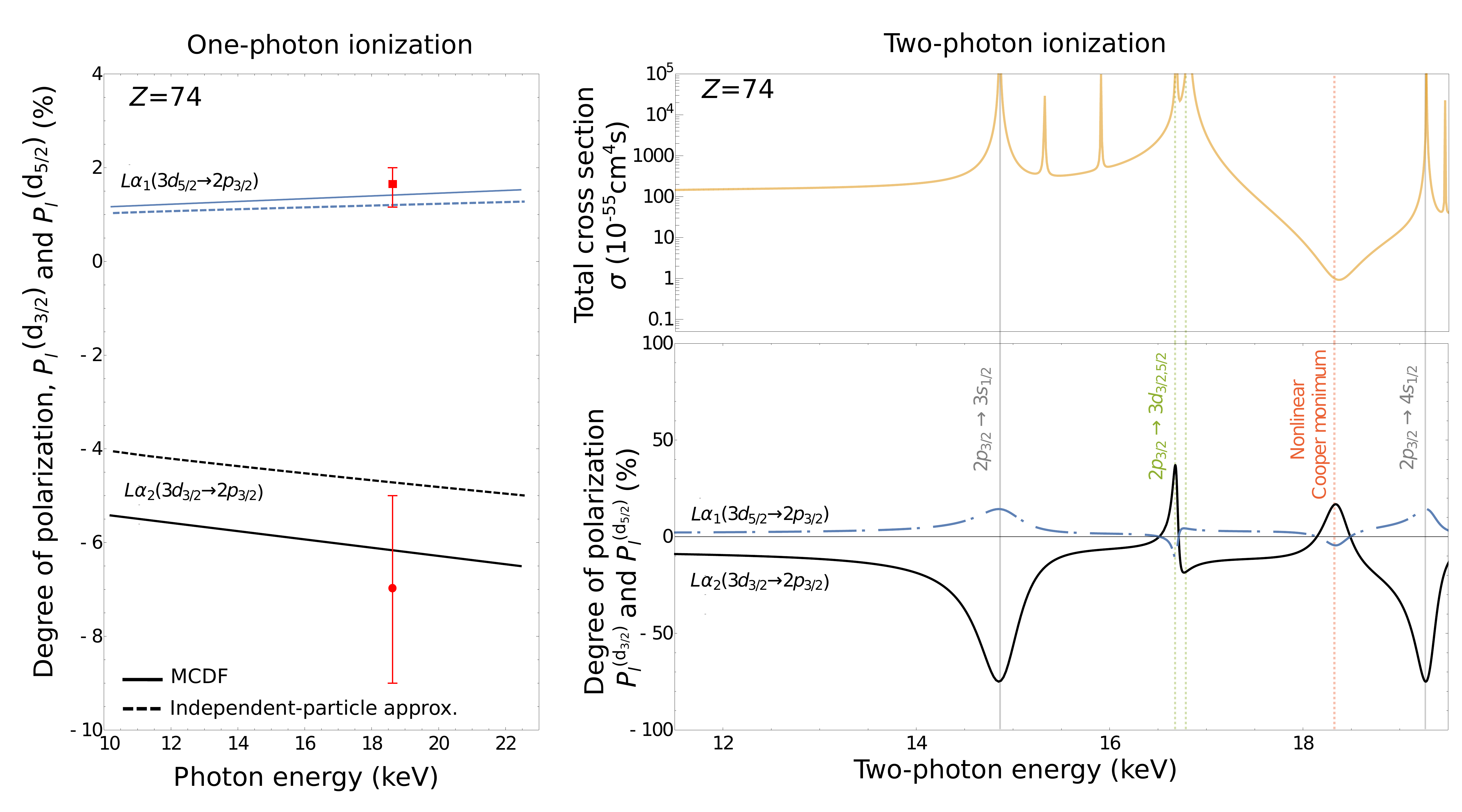}
    \caption{Degree of linear polarization of the characteristic $L\alpha$ lines following one- or two-photon ionization. Left: The degree of polarization $P_l^{(d_{3/2})}$ and $P_l^{(d_{5/2})}$ of $L\alpha_1$ and $L\alpha_2$ fluorescence photons, respectively as calculated by MCDF \cite{Kampfer:2016:033409} (full) or independent-particle approximation (dashed) approaches. Red markers represent the experimentally determined values \cite{Kampfer:2016:033409}. Right: Total cross section for the two-photon ionization process (top). The degree of polarization of $L\alpha_1$ and $L\alpha_2$ fluorescence at $\theta=0$ as calculated within independent-particle approximation. Vertical lines highlight local minima (maxima) of the degree of polarization. The right side of the figure shows, that transfer of linear polarization from incident to fluorescence photon is enhanced not only at photon energies matching an intermediate resonance, but also at the nonlinear Cooper minimum. }
    \label{Fig.MainFigure}
\end{figure*}

\section{Results and Discussion}
\label{Sec.Results}

Recently, K\"ampfer \textit{et al.} \cite{Kampfer:2016:033409} used unpolarized photons produced in an x-ray tube to ionize the $2p_{3/2}$ electron of neutral tungsten atom. The inner-shell hole was subsequently filled by an electron from $3d$ shell, and the characteristic $L\alpha_1$ and $L\alpha_2$ fluorescence photons were emitted. In their work, degree of linear polarization of these photons emitted perpendicular to the incident photon propagation was measured and reported together with theoretical calculation employing the multi-configuration Dirac-Fock method \cite{Jonsson:2013:2197, Fritzsche:2019:1}. Left side of Fig. \ref{Fig.MainFigure} presents the values from \cite{Kampfer:2016:033409} together with our calculations performed within independent-particle approximation. The good agreement of our prediction with experimental results as well as many-electron calculations presented in \cite{Kampfer:2016:033409} hints that our methodology may be further used to study x-ray fluorescence polarization phenomena in the regime of nonlinear interactions. 

A similar experiment as in Ref. \cite{Kampfer:2016:033409} is suggested here but for a two-photon ionization of a $2p_{3/2}$ electron of tungsten. For such a nonlinear inner-shell ionization, x-ray beams of sufficient high intensity are needed as produced, for instance, by XFEL. Since most free-electron laser beams are linearly polarized, we will consider two-photon ionization of $2p_{3/2}$ electron of neutral tungsten atom by linearly polarized photons. Generally, polarization transfer is maximized along the incident beam propagation, for this reason, we consider the degree of linear polarization of the fluorescence photons to be measured at $\theta=0$. In an experiment, it is impractical to perform the detection on the beam axis. For such cases, the degree of the photon polarization at an appropriate detection angle can be calculated using Eq. (\ref{Eq.StokesGeneral}). 

The right side of Fig. \ref{Fig.MainFigure} shows the degree of linear polarization (in percentage) of the $L\alpha$ fluorescence photons following two-photon ionization of a  $2p_{3/2}$ electron of tungsten by linearly polarized light (bottom), together with the corresponding two-photon ionization cross section (top) to guide the eye. The horizontal axis shows the combined energy of the two photons, i.e. the energy transferred to the atomic system. Already from the first glance, it is apparent that ionizing the neutral tungsten atom in nonlinear instead of linear regime results in high sensitivity of the fluorescence photon (or ion) polarization degree on the incident photon energy. Even more intriguing are the physical origins of each of the local maxima and minima in the signal. At low photon energies, in the nonresonant region, the degree of linear polarization is comparable to the case of one-photon ionization, however at higher energies, this value changes dramatically due to intermediate level resonances and the nonlinear Cooper minimum. The first and last local maxima (and minima) arise from the incident photon energy matching an intermediate $2p_{3/2} \rightarrow ns$ resonance (7.43~keV and 9.6~keV, respectively), they have been marked by solid gray lines. The two maxima marked with green dashed vertical lines originate from resonant transitions to the two fine-structure $2p_{3/2} \rightarrow 3d_{3/2,5/2}$ electron levels (8.3~keV, 8.4~keV). In tungsten atom, the $3s$ and $3d$ orbitals are of course occupied and hence, the $2p_{3/2}$ electron cannot be promoted into them. However, the resonance behavior arises from two-photon absorption, in which first photon ionizes the $3s$ or $3d$ electron and the second is excited the $2p_{3/2}$ electron into the corresponding hole, as experimentally verified in singly charged neon \cite{Kanter:2011:233001}. As the lifetime of the core-hole states are in the order of tens or hundreds of attoseconds, the ionization in this case proceeds non-sequentially only. For this reason, the lifetime of the atom in a state with $3s$ and $3d$ holes has been neglected. The last maximum, marked with red dashed line, originates from the significantly reduced contributions of the $p_{3/2} \rightarrow d_{3/2,5/2}$ ionization channels, which corresponds to the nonlinear Cooper minimum \cite{Hofbrucker:2019:011401}. 

The increased polarization transfer for incident photon energies matching an intermediate $s$ resonance and the nonlinear Cooper minimum have opposite signs, although they both arise due to relative increase of the $p \rightarrow s \rightarrow p $ channel contributions. This is the case because the individual fine-structure channels with $nd_{3/2, 5/2}$ intermediate states pass through their corresponding nonlinear Cooper minima at slightly different energies. As a consequence, the marked nonlinear Cooper minimum corresponds only to zero contributions of the $p_{3/2} \rightarrow d_{5/2} \rightarrow p/f$ channels, while the magnitude of the contributions of $p_{3/2} \rightarrow d_{3/2} \rightarrow p/f$ channels is comparable to $p_{3/2} \rightarrow s_{1/2} \rightarrow p$. However, since the channels with higher angular momentum generally dominate \cite{Fano:1985:617}, the polarization at the nonlinear Cooper minimum is still dominantly determined by the $d_{3/2}$ channel, and hence posses the same sign as the polarization signal at the corresponding resonance. One can also notice, that the resonances in the degree of polarization of the fluorescence photons are slightly shifted with respect to the corresponding resonances in the total cross sections. The shift of resonances of the total cross section with respect to other observables appears already for autoionization resonances in one-photon ionization, where the shift properties have been discussed in detail e.g. in Ref. \cite{Grum:2005:2545}.

The comparison of the left and right plots of Fig. \ref{Fig.MainFigure} demonstrates just how entirely new behavior of the light-matter interaction can be achieved with nonlinear processes. While the left-hand side figure contains only information about the initial and final electron states, two-photon ionization allows us to probe the electron complete structure of atoms, while utilize already available methods previously used in one-photon ionization case to obtain this information. Moreover, measurements of the fluorescence polarization can be carried out in order to determine the position of the nonlinear Cooper minimum, which would allow us to put the theoretical representation of the electron spectra (including positive and negative continua) to a test.


\section{Experimental consideration}
\label{Sec.Experiment}

Two-photon ionization experiments are typically performed by irradiating a sample with tightly focused XFEL (down to sub $\mu m$ size) beams and with the incidence x-ray energy being set below single-photon ionization threshold. Signature of the two-photon ionization mechanism is then confirmed by detection of subsequent x-ray emission signal resulting from relaxation of the produced ion \cite{Tamasaku:2014:313, Kayser:2019ky}. Dedicated experimental procedures are implemented to avoid higher harmonic components of the incident beam (monochromators) as well as energy jitter associated with x-ray generation at XFELs through the self-amplified spontaneous emission process \cite{Vinko:2012gx, Szlachetko:2016:33292}. We should note, that the nonlinear interactions of x-rays with a sample are not triggered selectively when using ultra-intense and ultra-short x-ray beams, and the linear contributions often surpass the nonlinear processes. As a consequence, the nonlinear x-ray signals are often accompanied with linear interaction contributions, which requires application of dedicated high energy resolution setups \cite{ Kayser:2019ky, Szlachetko:2016:33292} to resolve the linear and nonlinear contributions to the measured signal. 

To address the influence of linear ionization on the predicted results explicitly, let us shortly discuss it together with other effects such as the pulse length and time structure. As the cross-sections for two-photon ionization mechanisms are very low, the linear interaction of x rays with matter will surpass the nonlinear interactions. At the considered x-ray energies of 6~keV~-~9~keV, the dominant interaction will consist of single photoionization of $M$-shell electrons with corresponding binding energies spanning from 1.809~keV for $3d_{5/2}$ subshell to the energy of 2.820~keV for the $3s$ subshell and photoionization cross-sections in the order of $10^{-20}$ cm$^2$ \cite{NIST}. The core-holes created by single-photon ionization will decay mostly via Auger process and will thus increase charge state by one for every single Auger decay. The Auger cascades continue until the initial hole is transferred to the upmost subshell. It is safe to assume that single core-hole ionization may lead to the increase of atomic charge up to few hole states \cite{Rudek:2012:858, Rudek:2018:4200, Kayser:2019ky}. This high charge state will, however, be very quickly depopulated by electron transfer from surrounding atoms \cite{Erk:2013:053003, Rudenko:2017:129, Yoneda:2014:5080}. We calculated the two-photon ionization cross-sections assuming depletion of $P-$ and $O-$shells of tungsten atoms and found that the increase in charge state does not influence significantly the two-photon ionization cross-section values, nor the positions of resonances. We should note here, the expected change in binding energies of inner shell electrons as well as the position shift of the Cooper minimum (around $\approx 1$~eV) will be within energy bandwidth of the XFEL beam, and therefore will not affect the estimated two-photon ionization rates. Since the energy of the inner-shell levels does not shift significantly, the subsequent $L\alpha$ x-ray emission will also not be strongly influenced by outer-shell vacancies. This effect may be understood on the basis of strong and uniform screening of the $3d$ and $2p$ states by the remaining electron charge. The energy stability of inner-shell electrons was experimentally observed for $L\alpha$ x-ray emission for different oxidation states \cite{Wach:2020:689, Tirasoglu:2003:231}. It is reported that even $N$-satellites lines are reported to be within the linewidth of the $L\alpha$ x-ray emission line \cite{Czarnota:2013:052505} and only $M$-shell satellites lines will be outside the $L\alpha$ emission linewidth. However, appearance of $M$-satellites lines would indicate presence of three-photon absorption process within the $M$ core-hole lifetime which, according to the power law of $n$th order process $10^{-33}$cm$^{2n}s^{n-1}$ \cite{Faisal:2011}, will be of much lower probability than the two-photon absorption.

Up to now, cross-section values for two-photon ionization at hard x-ray energies have been determined for various gaseous and solid samples in a diverse range of incidence x-ray energies (see \cite{Tyrala:2019:052509} and references therein). The cross-section values are reported in the range of $10^{-53}$ cm$^{4}$s for two-photon $K$-shell ionization of Ne, $10^{-54}$ cm$^{4}$s for Cu and $10^{-58}$cm$^{4}$s for Ge and Zr. The determined $Z$-dependence of the total two-photon ionization cross section has been established to be between $Z^{-4}$ and $Z^{-6}$ \cite{Tyrala:2019:052509}. The main differences between reported cross-section values and $Z$-dependence originate from employed incidence x-ray energy and/or x-ray pulse length as well as its time structure. Indeed, the effects of second-order coherence and time structure to the determined cross section values are still discussed and the eventual influence have not been validated with experimental verification \cite{Sytcheva:2012:023414, Gorobtsov:2018:4498}. We should also stress that the time-independent theory employed for cross-section calculations implies the process of simultaneous absorption of both x-rays. The simultaneous absorption mechanism is also maintained in the case of $2p\rightarrow 3s \rightarrow e^-$ and $2p\rightarrow 3d \rightarrow e^-$  resonant two-photon ionization at around two-photon energy of 15~keV and 17~keV. It is assumed that first photo-absorption event leads to depletion of $3s$ or $3d$ core-level, which opens a strong dipole excitation channel for $2p$ core-electron into the partially empty $3s/3d$ levels, and hence, unveils the hidden resonance as explored experimentally by \cite{Kanter:2011:233001}. The core-hole lifetime for the $3s$ hole level is 0.04~fs and the lifetime for the $3d$ hole level amounts to 0.38~fs \cite{Campbell:1995dg}. The typical XFEL beam pulse lengths are in the order of few tens of femtoseconds and are much longer than the intermediate states core-hole lifetimes, therefore, the assumption of simultaneous two-photon ionization mechanism with intermediate atomic $3s$ and $3d$ states remains valid. Moreover, the expected polarization effect resulting from the produced tungsten ion via two-photon ionization is mainly determined by the angular momentum properties of the corresponding subshells. 

\begin{figure}
    \includegraphics[scale=0.4]{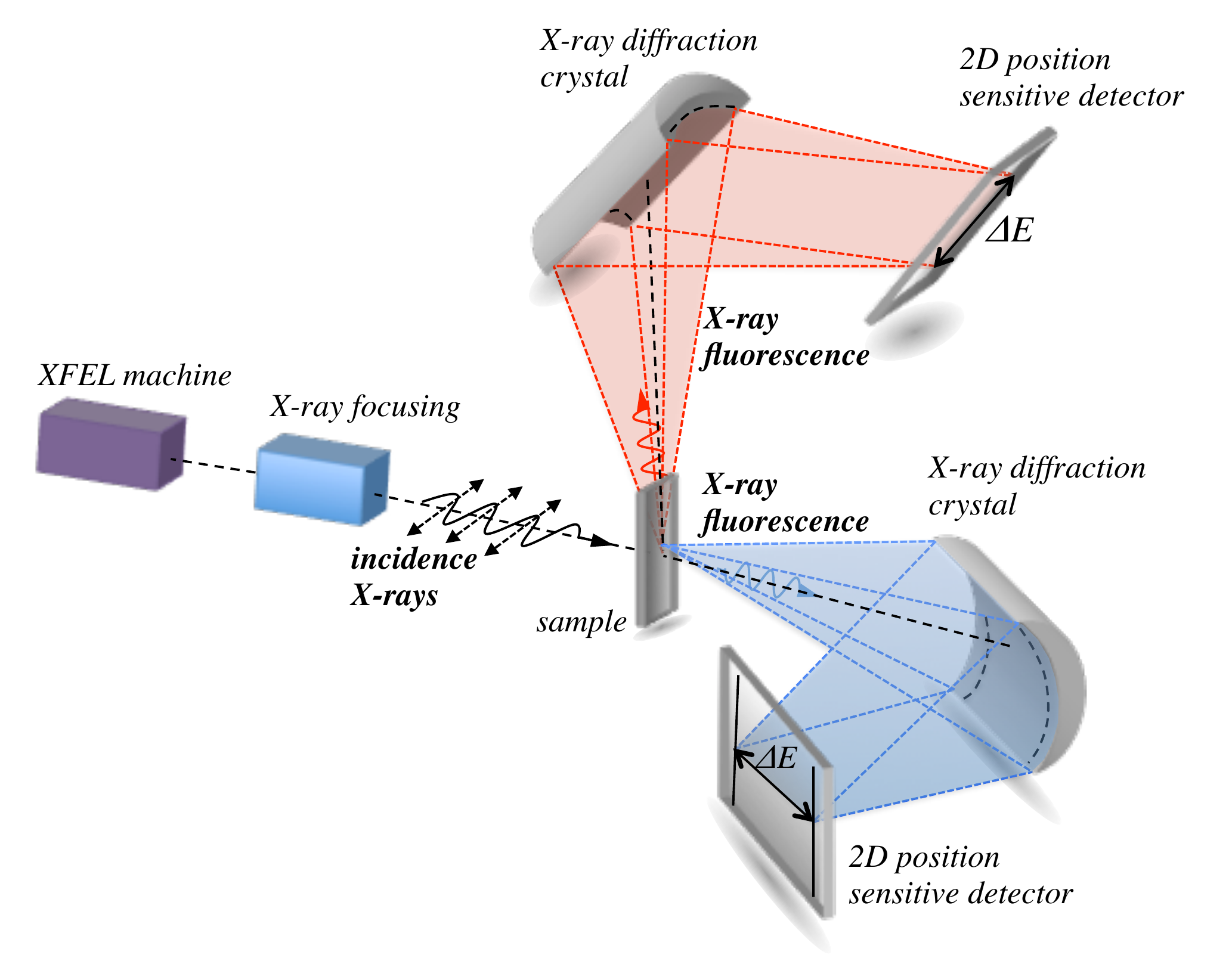}
    \caption{Schematic setup for polarization measurements of the x-ray fluorescence radiation following a two-photon ionization process. The setup explores polarization nature of XFEL beams and possibility of x-ray fluorescence analysis with dispersive spectrometer setups in parallel and perpendicular directions in respect to incidence x-ray beam.}
\label{Fig.ExpSetup} 
\end{figure}

Tungsten atoms ionized through the two-photon absorption process will decay to the ground state by electronic transition from higher electronic orbital. In the present case, we focus explicitly on the $L\alpha_{1}$ ($3d_{5/2}\rightarrow 2p_{3/2}$) and $L\alpha_{2}$ (3$d_{5/2} \rightarrow 2p_{3/2}$) decay channels. As the $3d_{5/2}$ and $3d_{3/2}$ final states are separated by 63~eV, a high energy resolution detection schemes are necessary to precisely measure the two x-ray emission lines. Polarization sensitivity of the experiment will be ensured by horizontal polarization of the XFEL beam in combination with dedicated spatial arrangement of the x-ray emission spectrometers. The x-ray spectrometer will be operated with Si(333) diffraction crystal giving 44.93$^\circ$ Bragg angle for $L\alpha_{1}$ and 45.36$^\circ$ for $L\alpha_{2}$ x-ray emission. Operation of the spectrometer at Bragg angles close to 45$^\circ$ diffraction geometry will provide strong rejection of fluorescence polarized out of the plane of the crystal surface. Typically, the rejection factor of 1000 is expected close to 45$^\circ$/45$^\circ$ diffraction geometry \cite{Lindle:1988dj, Kampfer:2016:033409}. Moreover, the diffraction spectrometer will ensure energy resolution of around 1-2~eV, which is smaller than the natural line widths of $L\alpha_{1,2}$ transitions of 6.5~eV \cite{Campbell:1995dg}, and large enough to precisely measure the shape of the x-ray emission. In order to achieve specificity on polarization of x-ray fluorescence, a two-spectrometer setup should be employed, providing measurements of x-ray emission spectra in parallel and perpendicular directions with respect to propagation of the XFEL beam. Schematic layout of the proposed experimental setup is drawn in Fig.~\ref{Fig.ExpSetup}. 

The experimental feasibility may be evaluated from the calculated cross sections for two-photon ionization, parameters of x-ray laser beams and characteristics of the x-ray spectrometer setups. x-ray fluorescence rate at detector of the x-ray spectrometer is given by $R=\Omega \sigma F^2 Y \Gamma$, where $\Omega$ is the spectrometer efficiency, $\sigma$ is the two-photon ionization cross-sections, $F$ is the incidence x-ray beam flux, $Y$ is the x-ray fluorescence yield for $L$ shell transitions and $\Gamma$ is the electron decay branching ratio. The efficiency of dispersive-type spectrometer employing cylindrically bent crystal and 2D position sensitive detector is in the range of 10$^{-4}$/eV \cite{Szlachetko:2012jx}. For the XFEL pulse, we assumed intensity of 10$^{11}$ photons/pulse, 30~fs pulse duration and focus of the x-ray beam down to 5$\times$5 $\mu m^2$, which provides the photon flux of around $10^{31}$ photons /(s cm$^2$) \cite{2013PhRvL.111d3001T, Doumy:2011:083002, Vinko:2012gx, Tamasaku:2014:313}. Using the two-photon ionization cross-section value of $10^{-55}$ cm$^4$s at nonlinear Cooper minimum and fluorescence yield of 28\% the x-ray rates at detector are estimated to be 280~photons/(s eV). This number confirms the experimental feasibility of detection polarization components of x-ray fluorescence induced by two-photon ionization process within a sub \% precision and few hundreds of seconds of acquisition time.

With the present capabilities of XFELs, the maximum {x-ray} flux is achieved when the machine is operating with Self Amplified Spontaneous Emission mode \cite{Andruszkow:2000:3825, Decking:2020:391}. For optimal experiments, short attosecond highly monochromatized x-ray pulses would be desirable in order to maximize nonlinear interaction signals with respect to linear ionization. This, however, is not presently possible, as hard x-ray FEL beam monochromatization via double crystal monochromators or by means of self-seeding modes or short-pulse operation will in turn deliver lower x-ray fluence \cite{Amann:2012:693}. In addition, such beam manipulation will broaden the temporal structure of the x-ray pulse due to the penetration of x-rays into monochromator crystals \cite{Shastri:2001:1131}. This imposes an experimental limit on exploring either short x-ray pulses or pulses with high monochromaticity. On the other hand, new methods are also developed that may allow us to overcome this physical limit \cite{Kayser:2019ky}. However, these methods have not yet been proven experimentally at sub-femtosecond time scales. Thus, while experimental studies of two-photon absorption processes with attosecond x-ray pulses would be of great interest, generation of isolated attosecond pulses in the soft x-ray range have been demonstrated only recently \cite{Duris:2020:30} and are not yet available for hard x-ray energies.

Finally, we would like to address sample preparation issues and eventual possibilities that may be explored during the proposed experiment. The XFEL beams are highly damaging due to their high power which leads to irreversible damage of the sample for every x-ray shot due to the heat dissipation occurring at picosecond time scales \cite{David:2011:57}. For this reason, the sample material should be refreshed on a shot-to-shot basis to avoid multiple x-ray hits in the same sample spot. In the case of tungsten, preparation of the sample in the gas phase may be very challenging at high enough densities and around the occupied sample environment at XFEL beamlines. Range of experiments \cite{Ghimire:2016:043418, Szlachetko:2016:33292, Tamasaku:2018:083901, Kayser:2019ky} explored solid state samples, where the sample is moving continuously to obtain fresh spot for each x-ray hit, however this is possible only at low (100~Hz) repetition rate machines and may be unlikely at XFELs such as the European XFEL \cite{Decking:2020:391} operating at MHz repetition rate. Another option is the use of tungsten-based compounds or molecules, because the slight change in the charge state of tungsten atoms does not influence two-photon ionization cross-sections and shifts the positions of resonances and the Cooper minimum only by a few eVs. Such complexes may be prepared in a liquid form and used in liquid-jet systems to ensure sample refreshment even at high repetition rate machines. We should note that comparison of the experimental results obtained for solid and molecular systems could allow us to address for example the effects of plasma creation in solid state samples on the measured x-ray signals.

\section{Conclusion}
\label{Sec.Conclusion}
In conclusion, we have studied the transfer of linear polarization from incident to fluorescence photons in the case of two-photon ionization of $2p_{3/2}$ inner-shell electrons. We have shown, that strong variation of the polarization transfer can be achieved not only at intermediate level resonances, but more importantly at the nonlinear Cooper minimum. We proposed an experimentally feasible scenario based on an already realized experiment which could be performed at number of current free-electron laser facilities. The results presented in this work, are however, general and hence could be demonstrated in two-photon ionization of other atomic systems. Measurements of the degree of linear polarization of fluorescence light could serve for the first detection of nonlinear Cooper minimum.

\begin{acknowledgments}
We acknowledge the support from the Bundesministerium f\"ur Bildung und Forschung (Grant No. 05K16SJA). This work was partially supported by the National Science Centre (Poland) under grant no. 2017/27/B/ST2/01890.
\end{acknowledgments}
\section*{appendix}
\label{Sec.Appendix}
The electric dipole amplitudes employed in Eq. (\ref{Eq.StatisticalTensors}) are
\begin{eqnarray}
U_{p_{3/2}}^2 &=& \frac{1}{98000 \pi^2}\left[ 25 \Rollt - 4\Rdtlt +9\Rdflt \right] \\ \nonumber
                &\times& \left[ 5\Rollt + \Rdtlt + 9\Rdflt \right],                \\
U_{p_{1/2}}^2 &=& \frac{1}{98000 \pi^2}\left[5 \Rolll + \Rdtll\right]^2,           \\
U_{f_{5/2}}^2 &=& \frac{1}{98000 \pi^2}\left[7\Rdttf + 3\Rdftf\right]^2,           \\
U_{f_{7/2}}^2 &=& \frac{1}{98000 \pi^2}\left[ \Rdfts \right]^2,                    \\\nonumber
\Bar{U}_{p_{3/2}}^{2} &=& \frac{1}{98000 \pi^2} \Big\{ 50 \Rollt \left[25 \Rollt + \Rdtlt \right.\\ 
                &+& \left. 54 \Rdflt \right] + 41\left[\Rdtlt\right]^2 + 54 \Rdflt \\\nonumber
                &\times& \left[7\Rdtlt + 39 \Rdflt\right] \Big\},
\end{eqnarray}
where the short notation for the transition amplitudes $U^{(l_{n{j_n}})}_{l_j} = U^{(l_{n{j_n}})}_{l_j}(1 1 1 1)$ was used. In general, these amplitudes are by 
\begin{equation}
    \label{Eq.TransitionAmplitudeRadial}
    U^{(l_{n{j_n}})}_{l_j}(p_1 J_1 p_2 J_2)=\SumInt_n\frac
        {R_{\kappa \kappa_n}(p_2J_2)R_{\kappa_n \kappa_a}(p_1 J_1)}
        {\epsilon_{n_a \kappa_a}+\omega-\epsilon_{n_n \kappa_n}}.
\end{equation}
The relativistic quantum number $\kappa$ is related to the total and orbital angular momentum quantum numbers as $\kappa = (-1)^{l+j+1/2} (j+1/2)$. In the transverse (velocity) gauge, these integrals are explicitly given for the magnetic  ($p=0$, or $pJ = M J$) transitions
\begin{eqnarray}
 R_{\kappa_f \kappa_i}(M J) &=& \int _ { 0 } ^ { \infty } d r \frac { \kappa _ { i } + \kappa _ { f } } { J + 1 } j _ { J } ( k r ) \big[ P _ { i } ( r ) Q _ { f } ( r )\nonumber \\ 
 &+& Q _ { i } ( r ) P _ { f } ( r ) \big], 
\end{eqnarray}
where the radial wave functions $P(r)$ and $Q(r)$ are obtained from single-electron Dirac equation, with a screening potential in the Hamiltonian, which partially accounts for inter-electronic interaction. We compared number of different potential models. The core-Hartree potential, which reproduces binding energies in a good agreement with experimental values, was used to produce the results presented in this work. For the electric transitions ($p=1$, or $pJ = E J$)
\begin{eqnarray}
 R_{\kappa_f \kappa_i}(E J) &=& 
 \int _ { 0 } ^ { \infty } d r \Bigg\{ - \frac { \kappa _ { i } - \kappa _ { f } } { J + 1 } 
 \big[ j _ { J } ^ { \prime } ( k r ) + \frac { j _ { J } ( k r ) } { k r } \big]  \nonumber\\
 &\times&\big[  P _ { i } ( r ) Q _ { f } ( r ) + Q _ { i } ( r ) P _ { f } ( r ) \big]  \\\nonumber  &+& { J \frac { j _ { J } ( k r ) } { k r } \big[ P _ { i } ( r ) Q _ { f } ( r ) - Q _ { i } ( r ) P _ { f } ( r ) \big] \Bigg\} } .
\end{eqnarray}
In the length gauge, this integral is given by
\begin{eqnarray}
  R_{\kappa_f \kappa_i}(E J) &=& 
  \int _ { 0 } ^ {\infty} dr  ~ j _ { J } ( k r ) 
  \big[ P _ { i } ( r ) P _ { f } ( r ) + Q _ { i } ( r ) Q _ { f } ( r ) \big] \nonumber\\  
  &+& j _ { J + 1 } ( k r ) \Big\{ \frac { \kappa _ { i } - \kappa _ { f } } { J + 1 } \big[ P _ { i } ( r ) Q _ { f } ( r )  \\ \nonumber
  &+& Q _ { i } ( r ) P _ { f } ( r ) \big] + \big[ P _ { i } ( r ) Q _ { f } ( r ) - Q _ { i } ( r ) P _ { f } ( r ) \big] \Big\}.
\end{eqnarray}
In the above expressions, $j_J(x)$ are the spherical Bessel functions. 
%
%

%
\end{document}